\documentclass[twocolumn]{aastex62}

\usepackage{graphicx}
\usepackage{epstopdf}
\usepackage{natbib}
\usepackage{amssymb}
\usepackage{amsmath}
\usepackage{float}
\usepackage{chngcntr}

\newcommand{\beq}{\begin{equation}}
\newcommand{\eeq}{\end{equation}}
\newcommand{\ida}{6045466571386703360}

\newcommand{\idc}{6045478558624847488}
\newcommand{\idz}{6045489283174903168}
\newcommand{\vara}{mmRR 1}
\newcommand{\blenda}{Blend 1}
\newcommand{\varc}{mmRR 2}
\newcommand{\othervar}{G3168}
\newcommand{\smallerid}{6045466571377393792}
\newcommand{\pera}{0.33190704}

\newcommand{\perc}{0.31673414}
\newcommand{\ampa}{1.0}  \newcommand{\ampc}{0.3}

\received{Nov. 1, 2018}
\revised{Dec. 13, 2018}
\accepted{Dec. 17, 2018}
\submitjournal{ApJ Letters}

\shorttitle{Ultralow-amplitude RR Lyrae Stars} 
\shortauthors{Wallace, Hartman, Bakos et al.}

\begin{document}

\title{Ultralow-amplitude RR Lyrae Stars in M4}

\author[0000-0001-6135-3086]{Joshua J. Wallace}
\affiliation{Department of Astrophysical Sciences, Princeton
  University, 4 Ivy Ln, Princeton, NJ 08544, USA}

\author[0000-0001-8732-6166]{Joel D. Hartman}
\affiliation{Department of Astrophysical Sciences, Princeton
  University, 4 Ivy Ln, Princeton, NJ 08544, USA}

\author[0000-0001-7204-6727]{G\'asp\'ar \'A. Bakos}
\altaffiliation{MTA Distinguished
Guest Fellow, Konkoly Observatory}
\affiliation{Department of Astrophysical Sciences, Princeton
  University, 4 Ivy Ln, Princeton, NJ 08544, USA}

\author[0000-0002-0628-0088]{Waqas Bhatti}
\affiliation{Department of Astrophysical Sciences, Princeton
  University, 4 Ivy Ln, Princeton, NJ 08544, USA}

\correspondingauthor{Joshua Wallace}
\email{joshuajw@princeton.edu}

\begin{abstract}
We report evidence for a new class of variable star, which we dub
millimagnitude RR Lyrae (mmRR).  From K2
observations of the globular cluster M4, we find that out of 24
horizontal branch stars  not previously known to be  RR Lyrae
variables, two show photometric variability with periods and shapes 
consistent 
with those of first overtone RR Lyrae variables.  The variability of
these two stars,
however, have amplitudes of only one part in a thousand, which is
${\sim}$200 times smaller than for any RR Lyrae 
variable in the cluster, and much smaller than any known RR Lyrae
variable generally. 
The periods and amplitudes are: \pera\ d with \ampa\ mmag
amplitude and \perc\ d
with \ampc\ mmag amplitude.
The stars lie just outside the instability strip, one blueward and one
redward.  The star redward of the instability strip also exhibits 
significant multi-periodic variability at lower frequencies.
We examine potential blend scenarios and argue that they are all
either physically implausible or highly improbable.
Stars such as these are likely to shed valuable light on many
aspects of stellar physics, including the mechanism(s) that set
amplitudes of RR Lyrae variables.
 \end{abstract}


\section{Introduction}
\label{sec:intro}

RR Lyrae stars are valuable astronomical tools.
They are used as standard candles, and 
to measure the helium abundance of stars in globular
clusters (GCs).
Space-based monitoring of RR Lyrae variables by
missions such as MOST \citep{walker2003}, CoRoT \citep{baglin1998},
and {\it Kepler}/K2 \citep{howell2014} has revealed 
new information on these objects.  For example, {\it Kepler} has revealed 
additional, low-amplitude oscillation modes in fundamental mode (RR0)
RR Lyrae variables \citep{molnar2012}, including RR Lyr
itself \citep{benko2010}. See \citet{molnar2018} for a more complete
list of these discoveries.

As part of continuing efforts to observe RR Lyrae stars, the GC
M4 (NGC 6121) was observed by {\it Kepler}/K2 in 2014 during its
Campaign 2 using a large superstamp that contained thousands of stars.
This and other K2 observations of GCs are the longest continuous
photometric surveys of populations of GC
stars,
monitored at the high precision that has been {\it Kepler}'s
hallmark.  
As part of our analysis of these data, we have discovered
two horizontal branch (HB) stars just outside the
instability strip that have photometric variations similar to first
overtone RR Lyrae 
(RR1) pulsators but with an amplitude ${\sim}$200 times lower than the
typical lowest amplitude RR1s.  We tentatively give these stars the name
``millimagnitude RR Lyrae'', or ``mmRR'' for short.  The two
stars are {\it Gaia} DR2 \ida\ (\vara) and
{\it Gaia} DR2 \idc\ (\varc). There is no previously identified variable
class that matches the properties of these stars, and if their
variability is associated with RR1 variability, then they would be by
far the lowest amplitude RR Lyrae variables yet 
discovered. Previous RR Lyrae searches 
would likely have been unable to find such low-amplitude objects, so
it is not  
surprising that they are only now being discovered by K2.

\section{Observations and Analysis}
\label{sec:observations}

\subsection{K2 Image Subtraction, Reduction, and Variable Search}
Our light curve extraction pipeline is very similar to
the image subtraction pipeline of \citet{soaresfurtado}.
Our specific pipeline, briefly described here, will receive a full
description in our publication of a catalog of M4 K2 variables.
We downloaded the 16
target pixel files (K2 IDs 200004370--200004385) 
that make up the M4 superstamp from the
Mikulski Archive for Space Telescopes and stitched them together
using \texttt{k2mosaic} \citep{barentsen2016}, producing a total
of 3856 images.  We removed images that were
blank or that otherwise would produce low
 quality photometry (usually due to excessive drift) and
 were left with 3724 images covering ${\sim}$78 days.
We reduced these images to a set of registered, subtracted images
using tools from the \texttt{FITSH} software package \citep{pal}.

We used the {\it Gaia} DR1
source catalog \citep{gaiamission,gaiadr1} as
both an astrometric \citep{gaiadr1astrometry} and photometric
\citep{gaiadr1photometry} reference catalog.  DR1 was used
instead of DR2 because our analysis began prior to 
DR2's release.  A conversion between {\it Gaia} magnitude $G$ and {\it Kepler}
magnitude K$_\mathrm{p}$ was determined, which, owing to the 
similar bandpasses of the two telescopes, was purely linear.
The converted $G$ magnitudes were used as reference magnitudes for
performing  image subtraction photometry 
on the subtracted images, using \texttt{fiphot} from \texttt{FITSH} and
a series of aperture sizes.  The 
aperture used for a given magnitude was determined by
calculating the RMS scatter of the final light curves
and finding the aperture that had the lowest median
RMS value in half-magnitude bins.

The light curves suffered from residual systematic variations due to
the roll of the spacecraft 
We performed a decorrelation of the measured
photometry against the
telescope roll using the process described by \citet{vanderburg2014}
and \citet{vanderburg2016}.
As part of the decorrelation, a B-spline was also
fit to the data with breakpoints set every 
1.5 days and removed from the data.  The 
\texttt{VARTOOLS} implementation \citep{vartools} of the 
trend filtering
algorithm \citep[TFA;][]{kovacs2005} was then used to further clean
up global trends in the final photometry.

Light curves were obtained for 4600 {\it Gaia} DR1 sources, 
which were searched for variability using the
Generalized Lomb-Scargle \citep[GLS;][]{lomb1976,scargle1982,zechmeister2009},
phase dispersion minimization \citep{stellingwerf1978}, box
least squares \citep{kovacs2002}, and auto-correlation
function \citep{mcquillan2013} methods as implemented in 
\texttt{astrobase} \citep{astrobase}. The results from these methods were 
searched by eye for significant variability.

\subsection{The Horizontal Branch Stars}

\begin{figure*}
\centering
\includegraphics{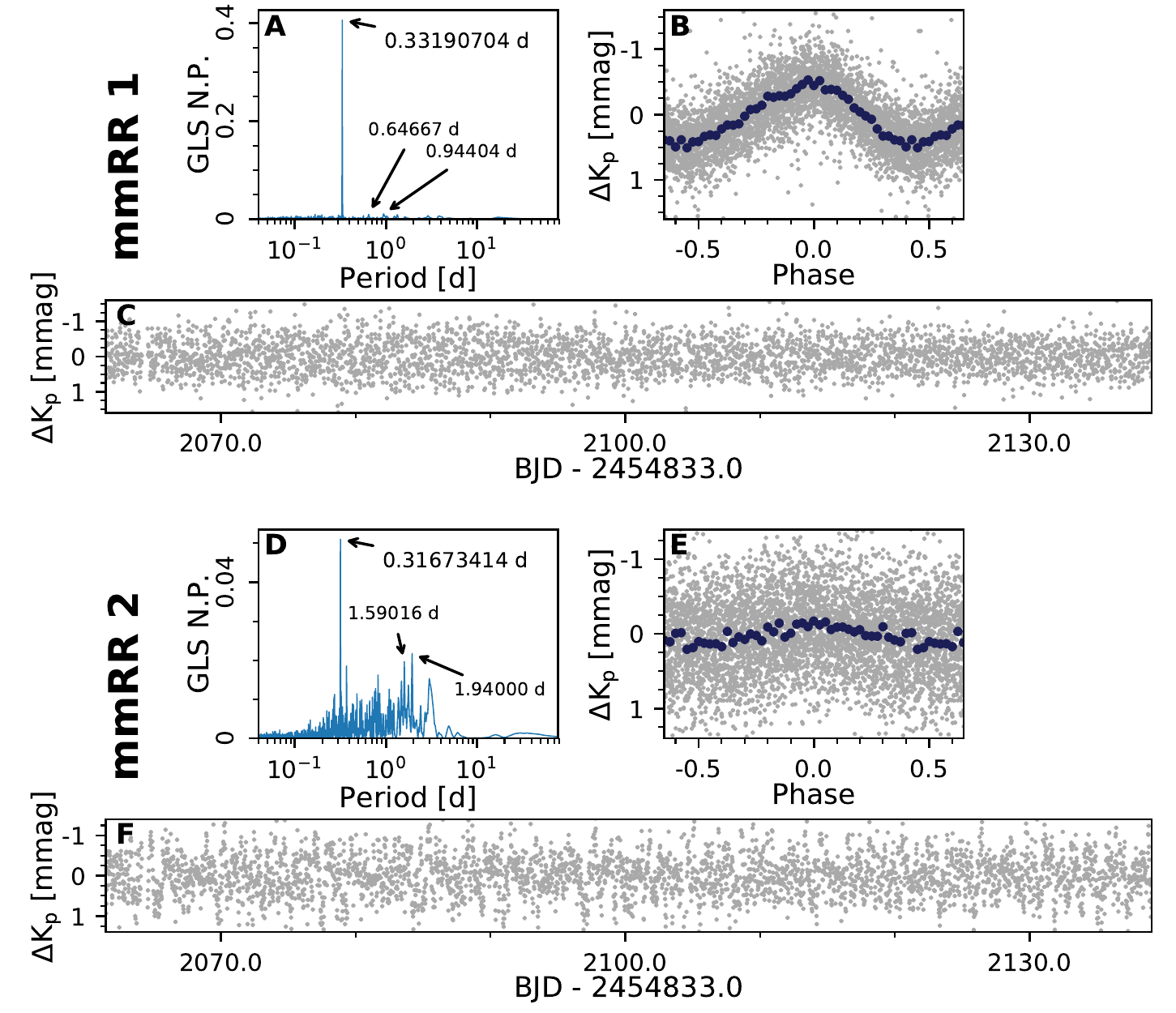}
\caption{
Light curves and periodograms for the two variable stars. 
Panels A, B, and C correspond to \vara\ and panels D, E, and F to \varc.
  Panels A and D
  show the GLS normalized power (N.P.) spectra with the three
  highest peaks in each case labeled.  Panels B and E
  show the phased light curves of each star, each folded at the
  GLS period with the highest peak. Panels C and F
  show the full light curves for each star.
  Gray points show individual measurements and blue points
  show binned-median values.
  All light curves have their median magnitudes subtracted off (\vara: 13.14153,
  \varc: 12.91269). We note
  possible notches in both phase-folded light curves just prior to maximum
  brightness. 
\label{fig1}}
\end{figure*}

\begin{deluxetable*}{lcccccc}
\label{tab1}
\tablecaption{Data on mmRRs}

\tablehead{\colhead{{\it Gaia} DR2 ID} & \colhead{$G$} &
  \colhead{R.A.} & \colhead{dec} & \colhead{period} &
  \colhead{amplitude} & \colhead{epoch}\\ 
\colhead{} & \colhead{(mag)} & \colhead{($\degr$)} &
\colhead{($\degr$)} & \colhead{(d)} & \colhead{(mmag)} & \colhead{(BJD-2454833.0)}} 

\startdata
\ida &  13.212 &  245.88458510 & -26.48151484 &  \pera &  \ampa & 2059.57\\
\idc &  13.047 &  245.89969745 & -26.43914199 &  \perc &  \ampc & 2059.47\\
\enddata

\tablecomments{Magnitude and position information from {\it Gaia} DR2.
  Epoch is the time of maximum brightness.}

\end{deluxetable*}

\begin{figure*}
\plotone{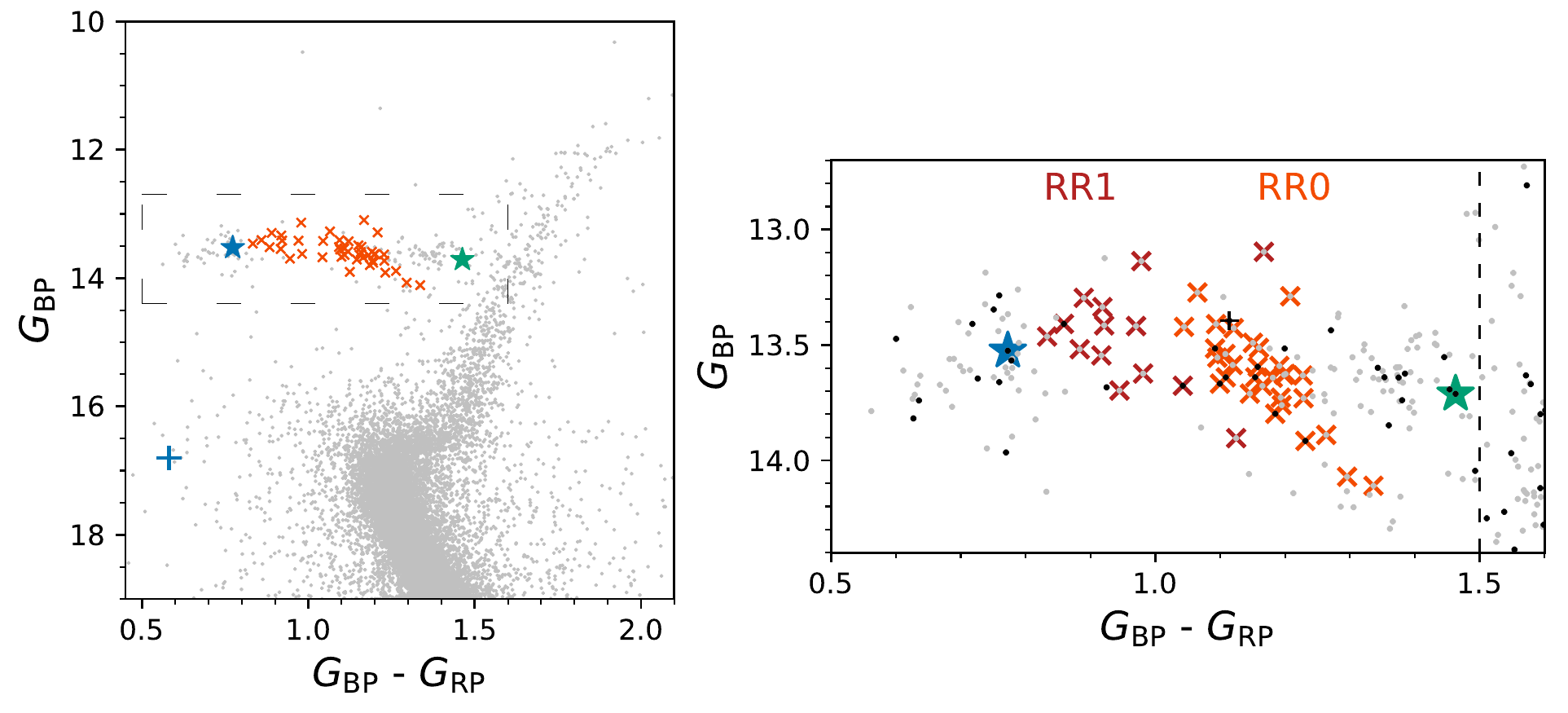}
\caption{{\it Gaia} DR2 CMDs for M4, with 
  $G_\mathrm{RP}$ and $G_\mathrm{BP}$ data
  taken from {\it Gaia} DR2 \citep{gaiadr2phot}.  Only objects with membership
  probabilities greater than 95\% are included.
  Left panel: red x's mark stars previously identified as RR Lyrae 
  in the catalog of \citet{clement2001}, June 2016 edition.
  The blue star marks \vara and the green
  \varc.  The blue cross marks a particular star blended
  with \vara\ (\blenda). Right panel:
  zoom-in of the portion of the left panel delineated by the dashed
  lines.  RR Lyrae variables are differentiated by subclass: light red for RR0
  and dark red for RR1, as indicated. For this panel, sources for
  which we have a K2 light curve  are marked in black instead
  of gray. Star \othervar\ is marked with a black cross. Our
  $G_\mathrm{BP}-G_\mathrm{RP} < 1.5$ cut for investigated 
  objects is shown with a vertical dashed line. \label{fig2}}
\end{figure*}

To determine cluster membership, we used {\it Gaia} DR2 proper motion
measurements \citep{gaiadr2astrometry} to determine cluster
membership. The proper motion of M4 ($\mu_{\alpha*}{=}-12.5$ mas/yr,
$\mu_{\delta}{=}-19.0$ mas/yr) is well separated from that of the 
field population.
We used
\texttt{scikit-learn} \citep{scikit-learn} to fit a two-component Gaussian
mixture model (GMM) to the proper motion measurements of all {\it Gaia} DR2
sources within 30\arcmin\ of the cluster center with reported
proper motions \citep[for full details, see][]{membershipcatalog2018}.

This Letter presents results of a variability search among  
the 34 HB stars for which we had light curves.  
HB stars were selected to be those with $14.3 < G_\mathrm{BP} < 13.0$
and $G_\mathrm{BP} - G_\mathrm{RP} < 1.5$ and a ${>}95$\% cluster
membership probability.
Of these, 10 were previously identified as RR
Lyrae variables \citep{clement2001}.  Of the other 24 HB stars, we
identified two low-amplitude variables with  ${\lesssim}$1 mmag amplitude sinusoidal variability and periods of
${\sim}$0.3 d and fell outside the locus of identified RR Lyrae stars.
Table~\ref{tab1} contains some information on these objects and
Figure~\ref{fig1} shows their light curves (full and phase-folded) and
associated GLS periodograms.  
 The light curves are published
online\footnote{\url{https://doi.org/10.5281/zenodo.2220532},
  \citet{mydata2018}}.  
The periods are consistent
with RR Lyrae variability, and the light curve shapes---in particular
the possible notches  
 just before maximum brightness for all two
stars---are similar to RR1.  The
amplitudes, however, are much smaller than any known RR1, which have
amplitudes of ${\sim}$200--350 mmag.  The variability search for
\vara\  detects only this 
sinusoid variability and its harmonics and aliases, while
\varc\ shows low-amplitude 
variability at a number of longer periods as well.  The positions
of these stars in the color-magnitude diagram (CMD) are 
shown in
Figure~\ref{fig2}.  Star \vara\ is blueward of the locus of RR
Lyrae stars and \varc\ is redward.  Several of the other stars 
redward of the full-amplitude RR Lyrae variables show
low-amplitude variability at multiple periods in the approximate range
0.3--5 d. 

We note a possible third star of interest, {\it Gaia} DR2 \idz\
(\othervar),  
 which has a ${\sim}$0.64-d sinusoidal period and
${\sim}$0.5 mmag amplitude and is in the locus of RR Lyrae
variables (marked in Figure~\ref{fig2} with a black cross).  We do not
 include it as an mmRR because it has stronger variability than \varc\
 at other periods (for example, a sinusoid variability at 1.67-d
 period of slightly smaller amplitude than the 0.64-d signal). We
 will further discuss this and the other M4 variables 
in a future work.

\subsection{Blend Scenarios}
\begin{figure}
\centering
\includegraphics[height=5.35in]{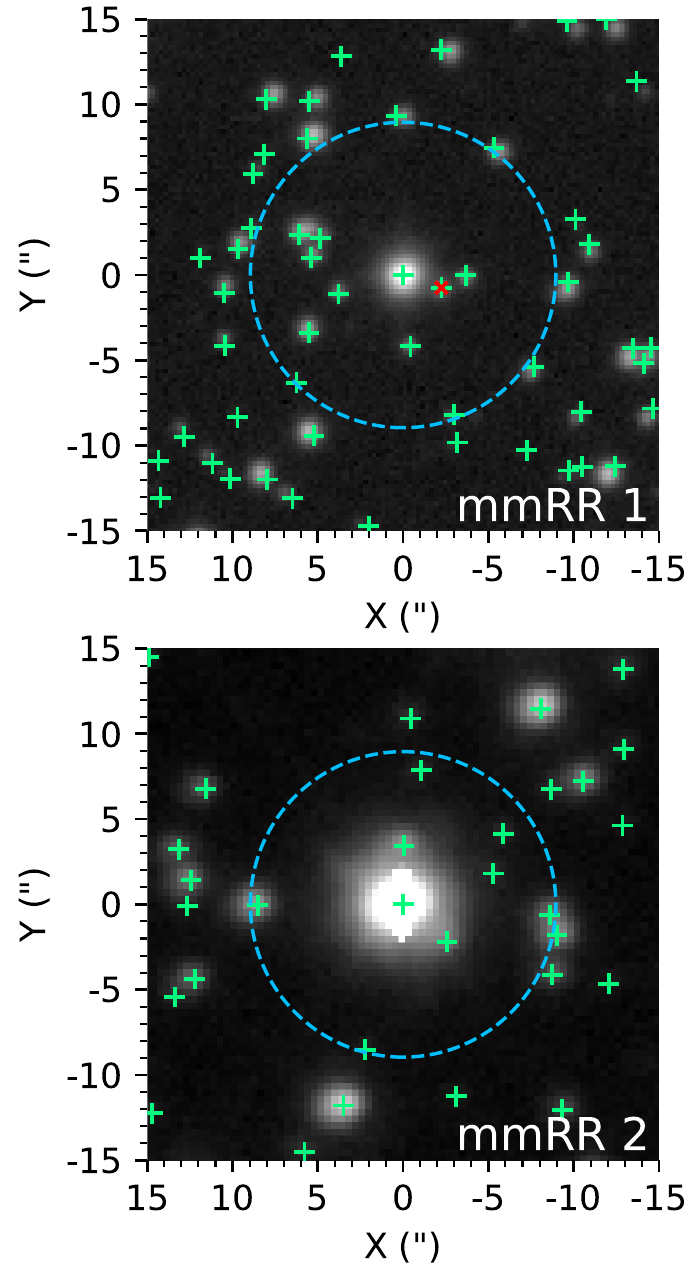}
\caption{Images of the two variable stars.  Crosses mark
  {\it Gaia} DR2 source positions. \blenda\ is additionally marked
  with a red x. The blue circles mark the size of aperture used for
  photometry extraction from the K2 data. The image for
  \vara\  is  
  from the M4 reference image of
  \citet{kaluzny2013a} and is rotated slightly relative to the {\it
    Gaia} source positions.    The image for  
  \varc\ (saturated in this image) was taken with a Sinistro detector
  on an LCOGT 1-m telescope 
  operated by Las Cumbres Observatory.
  The pixels in all three images are expressed in a logarithm
  scale. \label{fig3}}
\end{figure}

Figure~\ref{fig3} shows images of the two mmRRs, with
nearby {\it Gaia} DR2 sources marked.  The aperture used for
photometry extraction
is indicated, which has a radius of 2.25 {\it Kepler} pixels,
or ${\sim}$9\arcsec.  Both the apertures used and the individual
K2 pixels 
that these stars lie on are significantly blended.  We focused our blend
analysis on \vara\ due to its higher signal-to-noise ratio, but many
of our conclusions extend to \varc.

We searched for variability among the blended sources
by using an array of
0.51-pixel radius apertures on and around \vara\
to obtain focused photometry of the blended
objects from the K2 data.  This photometry underwent the same roll
decorrelation previously described but not TFA cleaning.  We
then searched for variability
at a period and flux amplitude matching the aperture centered on \vara.
Three apertures had a corresponding variability: the
one centered on \vara, and the two 
apertures located 0.51 pixels left and right roughly along the x-axis of the
image in Figure~\ref{fig3}.
We concluded the variability source could only be
\vara, {\it Gaia} DR2 \smallerid\ (\blenda, marked in
Figure~\ref{fig2} with a red x), or an unresolved blended source.

\blenda's location (blue cross in Figure~\ref{fig2})
in the 
CMD is unusual, particularly given its ${>}99$\% probability of cluster
membership.  The {\it Gaia} detector windows to measure
$G_\mathrm{BP}$ and $G_\mathrm{RP}$ 
are 2\arcsec.1 ${\times}$ 3\arcsec.5 \citep{arenou2018}, so it
is possible that the color measurements are 
significantly blended with \vara, perhaps inhomogeneously
between the two filters for it to appear bluer than
\vara.  It is also possible that \blenda\ is a
subdwarf B (sdB) star or a white dwarf (WD) blended with a main
sequence (MS) star.  We were unable to determine any physical MS-WD
combination that matched the measured color and magnitude for this
object. 
Many sdB stars are variable, but none in a way to match the
variability seen for this object (which, given the G$\approx$18
magnitude of this object, would need to have a ${\sim}$0.1 magnitude
amplitude).  All known types of sdB variables 
 are some combination of
too short of period, too small of amplitude, or too incoherent of
pulsations to explain the variability \citep[][chapter 12]{catelan2015}.
We also were unable to find any 
ellipsoidal variability of an sdB--MS binary that provided the necessary
variability amplitude (the highest unblended
amplitudes obtained were ${\sim}$0.01 mag).

If the color measurements are in error and this is an MS
star, the only variability scenarios that could match the
observed shape and period are rapid rotation of a heavily spotted star
or ellipsoidal
variability.  We were unable to find any physically plausible
ellipsoidal variability scenarios that matched the observed variability
and $G$-band magnitude. 
To estimate the probability of blending with a heavily spotted
fast rotator, we looked through the light curves for all
objects with $G{>}15$ and found five objects with periods
less than one day and sinusoidal
variability of roughly appropriate amplitude  when blended with an HB
object.  The search field was ${\sim}$149 
square arcminutes.  Since, from the aperture analysis, we know the
blend must be within about a {\it Kepler} pixel radius
(${\sim}$4\arcsec),   
the probability of one of these objects blending with \vara\ is
${\sim}5{\times}10^{-4}$. The probability
of finding two chance alignments out of 24 targets is very
small at ${\sim}7{\times}10^{-5}$.

Returning to \vara, 
the orbital separation needed for a ${\sim}$0.66 d binary orbit
including \vara\ is ${\sim}$3--4
R$_\sun$.  
{\it Gaia} DR2 \citep{gaiadr2apsis} measures the radius of \vara\ to be 2.8--4.1
R$_\sun$ (16th--84th percentiles).  We used 
\texttt{PHOEBE} \citep{phoebe1} to examine contact binary
scenarios and could find no physical scenario with the radius of
\vara\ being larger than ${\sim}$2.4 R$_\sun$, the Roche limit.  If
the radius of \vara\ is indeed exceptionally small to allow a contact
binary scenario, only companions with masses
between $0.08$\,M$_{\sun}$ (with a face-on orbit) and ${\sim}1$
Jupiter mass (with inclination ${\lesssim}45^{\circ}$) could produce
millimagnitude amplitudes. Given the even larger
radius that \vara\ would have had when on the red giant
branch, such a system would be a
post-common-envelope-binary. Approximately one third of WDs
are known to have short-period post-common-envelope binary
companions, with the majority having secondary stars of mass less than
$0.25$\,M$_{\sun}$ \citep{schreiber2010}.  While we are unaware
of estimates for the occurrence rate of such systems with HB primary
stars, the possibility of an HB star having a low-mass contact binary
companion cannot be dismissed out of hand. However, finding two of
these systems on low inclination orbits (which are less likely
than higher inclinations assuming random orientations), without also
finding systems on higher inclination (and thus higher photometric
amplitude) orbits is unlikely. Moreover, the inconsistency between the
measured stellar radius from {\it Gaia} and the upper limit on the
radius for a contact binary is strong evidence that this scenario does
not explain the observations.

\begin{figure}
\plotone{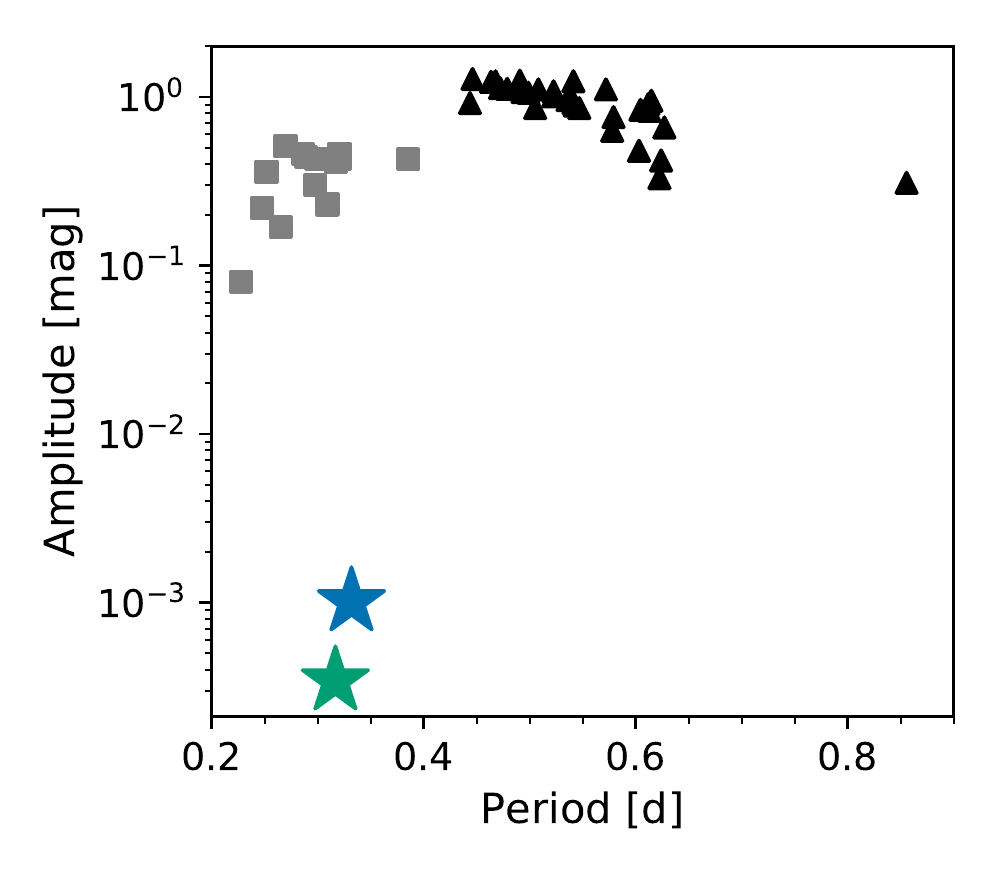}
\caption{Periods and amplitudes of the two mmRRs and
  the RR Lyrae 
  stars in M4.  As in Figure~\ref{fig2}, 
  the blue star marks \vara and the green
  marks \varc. RR0 are shown as black
  triangles and RR1 are shown as gray squares.  The data for RR Lyrae
  variables are
  from \citet{clement2001}.  The two new variables
  have much lower 
  amplitudes than any RR Lyrae star in the cluster.\label{fig4}}
\end{figure}

Finally, we consider an undetected
background RR1 or short-period Cepheid variable.  An RR1 would need to be
${\sim}$200--350 times dimmer than \vara\ to get a millimagnitude
blended amplitude.
With \vara\ having $G{=}13.23$, the background RR1 would
need to have $G{\approx}$18.9--19.6.   
We used the {\it Gaia} DR2
RR Lyrae variable catalog \citep{holl2018,clementini2018}
to determine the surface density of RR Lyrae variables with $G$
magnitudes in the appropriate range in the field near M4, 
finding ${\sim}$2 RR Lyrae variables per 0.7 square degrees.  
Mirroring our estimation of rapidly rotating spotted star blending, we
get a blend probability of ${\sim}6{\times}10^{-6}$.
The probability
of finding two chance alignments out of 24 targets is vanishingly
small at ${\sim}1{\times}10^{-8}$.  Even if the catalog of RR Lyrae we
are using has a completeness as low as 15\% as it does in the Galactic
Bulge \citep[][Table 3]{holl2018}, the probability of chance
alignment is still minuscule.
There are even fewer
background Cepheid variables ({\it Gaia} detected none the areas we
searched for RR Lyrae variables) and they typically have much longer
periods, so the probability of blending with a
background Cepheid is even smaller.

The signal-to-noise ratio for \varc\ was 
not high enough for our small aperture array to disentangle
specific possible sources of the variability.  We note, however, that
all of the {\it Gaia} DR2 sources within 5\arcsec\ of 
  \varc are proper 
motion members of the cluster.
Because of this, arguments similar to those for the possible blend
scenarios of \vara\ and \blenda\ prevail. 
We note that \varc\ has a relatively large radius in {\it Gaia} DR2
(7.8--8.3 R$_\sun$), which would make it impossible to host a binary
object at a ${\sim}$0.63-d period orbit.
We also checked that
the periods of the three variables do not match 
any previously identified RR Lyrae star in the cluster, nor do they
match any other variable signal found in our 
  light curves from the M4 superstamp.
{\it Gaia} detects no variables within 10\arcsec\ of the two mmRRs.

\section{Discussion}
\label{sec:discussion}
From the evidence presented, we conclude
that the most likely explanation for the observed variability 
is a previously unreported kind of stellar variability that,
based on the
locations in the CMD and variability periods and shapes, is  possibly related
to RR Lyrae variability.
Figure~\ref{fig4} shows the periods and amplitudes of the mmRRs
relative to the RR Lyrae variables in M4.  Their amplitudes (\vara: \ampa\
mmag,  \varc: \ampc\ mmag) are 
much lower than any previously observed RR Lyrae star, which have
amplitudes of ${\sim}$200 mmag and greater.

We note here \citet{buchler2005} and \citet{buchler2009},
who used data from the MACHO and OGLE databases to find ${\sim}$30
objects near the Cepheid instability strip of the 
LMC with
amplitudes ${\lesssim}$0.01 mag.  At least ${\sim}$20 of these objects
are members of the LMC.  
These match the predicted strange Cepheids of \citet{buchler1997}.
These mmRRs may be the very similar strange RR Lyrae predicted
by \citet{buchler2001}.  The amplitudes, shapes, and CMD locations match the
predictions, but the periods (which would be coming from the 8th-10th
radial overtones) are longer than predicted.

We also note once again \othervar, the possible
third mmRR we found, as well as the other HB stars redward of the
known RR Lyrae stars that had multi-periodic photometric variability of
periods of approximately 0.3--5 days.  These stars are perhaps
connected to the mmRRs and will be described more completely later.

If 
these objects do represent a new class of variability, why have no
similar objects been discovered previously?  As mentioned in
Section~\ref{sec:intro}, {\it Kepler}/K2 has enabled discovery of very small
amplitude modes in RR Lyrae variables, seemingly commonplace yet
undetected in over a century of observations of these stars.  The
mmRRs appear to share a similar story. 
We make particular mention of RR Lyr, an RR0, 
which has been shown by {\it Kepler} to have small amplitude first overtone
pulsations \citep{molnar2012}, a phenomenon perhaps connected to these
mmRRs.  Finally, theoretical work indicates
that convection and viscous damping are the likely physical process
that set the amplitudes of RR Lyrae
variables \citep{kollath1998,smolec2008,geroux2013}; mmRRs could be 
valuable in further developing this understanding.

\acknowledgements

We thank A. Vanderburg for assistance with the roll decorrelation and W. Pych
for providing CASE light curves that were useful 
in our initial vetting. G\'AB thanks J. Jurcsik, G. Kov\'acs, and
L. Moln\'ar for useful discussions while at Konkoly Observatory.
This research includes data collected by the K2 mission, funding for which
is provided by the NASA Science Mission directorate. 
The K2 data were obtained from the
Mikulski Archive for Space Telescopes (MAST). STScI is operated by the
Association of Universities for Research in Astronomy, Inc., under
NASA contract NAS5-26555. Support for MAST for non-HST data is
provided by the NASA Office of Space Science via grant NAG5-7584 and
by other grants and contracts. 
This research includes data from the European Space Agency (ESA)
mission {\it Gaia} (\url{https://www.cosmos.esa.int/gaia}), processed by
the {\it Gaia} Data Processing and Analysis Consortium (DPAC,
\url{https://www.cosmos.esa.int/web/gaia/dpac/consortium}). Funding
for the DPAC has been provided by national institutions, in particular
the institutions participating in the {\it Gaia} Multilateral
Agreement.
This research has made use of the SIMBAD database
(operated at CDS, Strasbourg, France),  NASA's Astrophysics Data
System Bibliographic Services, and observations from the LCOGT
network.

\facilities{du Pont (TEK5 2K), {\it Gaia}, {\it Kepler}, LCOGT (Sinistro)}

\software{\texttt{astrobase} \citep{astrobase}, 
  \texttt{astropy} \citep{astropy}, 
  \texttt{FITSH} \citep{pal}, 
  \texttt{k2mosaic} \citep{barentsen2016},
  \texttt{matplotlib} \citep{hunter2007},
  \texttt{numpy}  \citep{numpy},
  \texttt{PHOEBE 1.0} \citep{phoebe1},
  \texttt{scikit-learn} \citep{scikit-learn},
  \texttt{scipy} \citep{scipy}, 
  \texttt{VARTOOLS} \citep{vartools}
}

\bibliographystyle{aasjournal}

\end{document}